# Mass spectra of meson molecular states for heavy and light sectors

S. Rahmani[1*] and H. Hassanabadi[1]

[1]Physics Department, Shahrood University of Technology, P. O. Box 3619995161-316, Shahrood, Iran

* Email:s.rahmani120@gmail.com

**Abstract:** We obtain mass spectra of the light and heavy meson-antimeson (molecular states) sectors by using a nonrelativistic potential model with Coulomb and one pion exchange potential terms for meson-meson interaction. The digamma decay widths are also obtained for the light sector. We compare our results with available experimental and theoretical data.

## 1 Introduction

Recent experimental data indicate a number of new exotic states of mesons [1-3] which cannot be considered as simple $qqq$ and $q\bar{q}$ forms. They are suspected to have a molecular character. Meson-antimeson molecules are very interesting objects from the theoretical point of view and have been considered for a long time. Jaffe suggested that there should be hadronic resonances with $qq\bar{q}\bar{q}$ flavor quantum numbers and also that some of the observed hadronic resonances should be interpreted in this way [4]. Hanhart et al. analyzed two-photon decays of hadronic molecules [5]. Liu and Zhu studied $Y(3930)$ and $Y(4143)$ as heavy molecular candidates and concluded that both of them are very good molecular states composed of a pair of vector charm mesons [6]. Based on the meson exchange model, Liu et al. performed a systematic study of three types of possible heavy molecular states: PP, PV, VV including $D\bar{D}/B\bar{B}$, $D^*\bar{D}/B^*\bar{B}$ and $D^*\bar{D}^*/B^*\bar{B}^*$ respectively [7]. In 1976, Voloshin and Okun studied the interaction between a pair of charmed mesons and proposed the possibility of molecular states involving charmed quarks [8]. The charmonium-like resonance $X(3915)$ has been investigated as a molecular bound state by Li and Voloshin [9]. $Y(4143)$, $Y(3940)$ and $Y(4140)$ have also been investigated as molecular bound states [10-12]. Braaten et al. studied $XYZ$ mesons as bound states in Born-Oppenheimer potentials for a heavy quark and antiquark [13]. Brodsky et al. presented a dynamical picture to explain the nature of the exotic $XYZ$ states based on a diquark-antidiquark open-string configuration [14]. Based on a diquark-antidiquark model, the hidden charm tetraquark spectra and the decay widths of the hidden charm tetraquarks into two charmed mesons were studied by Zhu [15]. Our aim in this work is to study some of the properties of hadronic molecules in the light and heavy sectors, including mass, binding energy and digamma decay widths, by considering a molecular-like interaction. We use a pseudoscalar and a vector meson for the molecular meson-antimeson system. Thus three possible combinations are considered: pseudoscalar–pseudoscalar (PP state), pseudoscalar–vector (PV state) and vector–vector (VV state). In the next section, we first consider the Hamiltonian of the system including the Coulomb and One Pion

Exchange Potential (OPEP) and evaluate the wave function and energy of the meson-antimeson system. We add a spin-dependent interaction as the perturbative term. Then we evaluate the masses, binding energies and digamma decay widths of some heavy and light meson-antimeson systems. Section (3) includes our results and discussion. In the final section we present our conclusions.

**2 Formalism**

We consider the molecular system as a meson-meson bound state. *V(r)* includes the molecular interaction $V^0(r)$ with a pion exchange potential $V_\pi$ as

$$V(r) = V^0(r) + V_\pi + V_{SD} \qquad (1)$$

where $V_\pi$ is the OPEP. We study the meson-antimeson systems (hadronic molecules) which are molecular bound state systems. These loosely bound states are similar to the deuteron-like (proton–neutron) system. To study the mesons or multiquark spectra, different potentials like Coulombic, confinement, spin dependent and combinations of these have been applied. Each of the potentials should be capable of describing the bound state properties at short range and long distances. We do not include the confinement potential for the dimesonic systems. Dimesonic systems can be considered as molecular-type bound state systems. Therefore we need a molecular interaction-like Coulomb potential. Short-range nucleon-nucleon interactions can be considered to be a residual color force extending outside the boundary of the proton or neutron. Hence we consider the meson-antimeson interaction as the Coulomb potential:

$$V^0(r) = \frac{\beta}{r} + \gamma \qquad (2)$$

where $\beta = -\alpha_s$ [16] and $\gamma$ is a scale parameter of the potential. In fact $\gamma$ is a regularization dependent mass term. We obtained $\gamma = 0.09$ GeV, 0.07 GeV, 0.07 GeV, 0.05 GeV and 0 GeV in the case of the heavy charmed, heavy bottom, and light sectors for PP, light sector for PV, and light sector for VV dimeson states by fitting the mass of each of these systems with other theoretical or available experimental data. More explanation is given in Section (3). $\alpha_s$, the strong running coupling constant, is defined as [17,18]

$$\alpha_s(\tilde{M}^2) = \frac{4\pi}{(11 - \frac{2}{3}n_f) \ln \frac{\tilde{M}^2 + M_B^2}{\Lambda_Q^2}} \qquad (3)$$

with $\tilde{M} = 2\mu$ where $\mu = \frac{m_a m_b}{m_a + m_b}$ being the reduced mass of the di-mesonic system and $M_B = 1$ GeV. In Eq. (3), $\Lambda_Q$ is the QCD scale, taken as 0.413 GeV, and $n_f$ stands for the number of flavours. In our calculations we have taken the input masses of mesons as given in Table 1. We have tabulated values of the parameters used in our model for di-mesonic systems in Table 2. We use OPEP perturbatively for all meson-antimeson combinations. In the meson-antimeson case, the OPEP takes the form [19]

$$V_\pi = \frac{1}{3} \frac{g_8^2}{4\pi} \left( \frac{m_\pi^2}{4 m_a m_b} \right) (\tau_a . \tau_b)(\sigma_a . \sigma_b)$$
$$\times \left( \frac{e^{-m_\pi r}}{r} - (\frac{\Lambda_\pi}{m_\pi})^2 \frac{e^{-\Lambda_\pi r}}{r} \right) \qquad (4)$$

where $m_{\pi^0} = 0.134$ GeV and $\frac{g_8^2}{4\pi} = 0.67$ [19]. Also, the OPEP depends on the explicit cut of $\Lambda_\pi$. $\Lambda_\pi$ is the form factor. It appears due to the dressing of quarks and is assumed to be proportional to the exchange meson mass and the flavor independent parameter $\Lambda_0$ as [19]

$$\Lambda_\pi = k m_\pi + \Lambda_0, \qquad (5)$$

where $k = 0.81$ and $\Lambda_0 = 2.87 (fm^{-1})$ [19]. The values of the spin-isospin factor are taken as $(\tau_a \cdot \tau_b)(\sigma_a \cdot \sigma_b) = -3, 1$ for $I=0, 1$ in PV states. We assume the values $(\tau_a \cdot \tau_b)(\sigma_a \cdot \sigma_b) = -6, -3, 3$ when $I=0$ and spin $S=0,1,2$ while when $I=1$ and $S=0,1,2$, we have $(\tau_a \cdot \tau_b)(\sigma_a \cdot \sigma_b) = 2, 1, -1$ in the case of VV states. Now let us consider the spin-dependent interaction as [20]

$$V_{SD} = \frac{8}{9} \frac{\alpha_S}{m_a m_b} \vec{S}_1 \cdot \vec{S}_2 |\psi(0)|^2, \quad (6)$$

where the square of the wave function at the origin is

$$|\psi(0)|^2 = \frac{\mu}{2\pi} < \frac{dV^0(r)}{dr} >, \quad (7)$$

with $\mu$ the reduced mass of the di-mesonic system. One might expect that the effect of the spin dependent interaction and the one pion exchange potential have small corrections compared to the effect of the other terms in the Hamiltonian. In fact, spin splittings arise from additional terms in the Hamiltonian that can be treated as perturbations [13]. Therefore the spin dependent interaction [13,16,21] and one pion exchange potential [22] may be treated perturbatively. The Schrödinger equation is solved for the parent part (Coulomb potential). By considering the Coulomb potential and $\psi_{nl}(r) = \frac{u_{nl}(r)}{r}$, we can write

$$\frac{d^2 u_{n,l}}{dr^2} + \frac{1}{r^2}[-l(l+1) - 2\mu\beta r + 2\mu(E_{n,l} - \gamma)r^2] u_{n,l} = 0. \quad (8)$$

Then by using the NU (Nikiforov-Uvarov) method [23,24], the energy and wave function of the system become the following:

$$E_{n,l} = -\frac{\mu\beta^2}{2(n+l+1)^2} + \gamma, \quad (9)$$

$$\psi_{n,l}(r) = N_{n,l} r^{-\frac{1}{2} + \sqrt{\frac{1}{4} + l(l+1)}} e^{-\sqrt{-2\mu(E_{n,l} - \gamma)}r} \times L_n^{2\sqrt{\frac{1}{4} + l(l+1)}} (2\sqrt{-2\mu(E_{n,l} - \gamma)}r) \quad (10)$$

where $N_{n,l}$ is the normalization constant and $L_n^m(x)$ represents the Laguerre polynomial. We have calculated the S-wave state masses of the low-lying and heavy meson-antimeson states. The normalization constant is calculated as

$$N_{0,0} = \sqrt{\frac{(\beta^2 \mu^2)^{\frac{3}{2}}}{\pi}}, \quad (11)$$

for the $n=0$ and S-wave state. For the ground state of $\psi(0)$ we can write

$$\psi(0) = \frac{\sqrt{-\beta^3 \mu^3}}{\sqrt{\pi}}. \quad (12)$$

The mass of the meson-antimeson, $M$, including both low-lying and heavy states, is given by

$$M = m_a + m_b + E_{n,l} + <V_{SD}>_{n,l} + <V_\pi>_{n,l}. \quad (13)$$

where $<V_{SD}>_{n,l}, <V_\pi>_{n,l}$ can be defined as

$$<V_{SD}>_{n,l} = <\psi_{n,l}(r)|V_{SD}|\psi_{n,l}(r)>, \quad (14)$$

and

$$<V_\pi>_{n,l} = <\psi_{n,l}(r)|V_\pi|\psi_{n,l}(r)>, \quad (15)$$

respectively. By replacing Eq. (9) in Eq. (13) and using Eqs. (14), (15) for the case of $n=0$ and the S-wave state, we arrive at

$$M = m_a + m_b + \gamma - \frac{\mu\beta^2}{2} + <\psi_{0,0}(r)|V_{SD}|\psi_{0,0}(r)> + <\psi_{0,0}(r)|V_\pi|\psi_{0,0}(r)>, \quad (16)$$

where the spin dependent interaction and OPEP are added separately. To calculate the spectra of the excited states, we can obtain the wave function and energy of the system for other values of $n$. For example, one can account for the exited mass spectra in the case of $n=1,2$ and so on. $m_a$ and $m_b$ are the masses of the first meson and second meson respectively in dimesonic systems. For example, in the dimesonic molecular system $D - \bar{D}_s$, $m_a$ is the mass of meson $D$ and $m_b$ is the mass of meson $D_s$, which are $m_a = 1.8696 GeV$ and $m_b = 1.9683 GeV$ respectively, as listed in Table 1. The binding energy is

$$B.E. = E_{0,0} + <V_{SD}>_{0,0} + <V_\pi>_{0,0}, \quad (17)$$

for the $n=0$ and S-wave state. The two-photon decay for the light sector of exotic states like $a_0(980)$, $f_0(980)$, $f'_2(1525), f_2(1565), a_2(1700)$ has been observed experimentally. Studies of the two-photon decay of scalars could distinguish among different scenarios for scalar meson structure. In this field, light scalar mesons $a_0(980)$ and $f_0(980)$ are the most studied. Predictions of various models for these cases are different within the molecular model for scalars. They vary from 0.2 keV in Ref. [25] to 0.6 keV in Ref. [26], to 6 keV in Ref. [27], to 0.285 keV [16] and to $(0.22 \pm 0.07)$ keV [5]. Using the wave function at the origin and masses of the dimesonic systems, we can obtain the digamma decay width as follows [26,27]

$$\Gamma_{\gamma\gamma} = \frac{\pi \alpha^2}{M^2} |\psi(0)|^2, \quad (18)$$

where $\alpha = \frac{e^2}{4\pi} \approx (137)^{-1}$ is the fine-structure constant.

## 3 Results and Discussion

The input masses of Table 1 are taken from Ref. [28]. In Table 3 we have reported our calculated masses of some *D-D* bound states and compared them with Refs. [1,29,28,30,31]. Our calculations for bottom meson-antimeson systems are shown in the third column of Table 4. Table 5 shows our results for the case of PP states in the light sector. In Table 6 we show the masses of the light meson-antimeson sector for PV states. In Table 7 we report our results for VV states for the light meson-antimeson sector. We have calculated the excited spectra of heavy charmed meson-antimeson states for n=1 in Table 8. In Table 3, we get $\gamma = 0.09$ GeV by using the value of the mass of $D_s - \bar{D}^*$ ($\frac{1}{2}(1^{+-})$), 3.969 GeV [30]. By using $M^{B^*-\bar{B}^*(0^+(2^{++}))} = 10.604 GeV$ [32], $\gamma$ becomes 0.07 GeV in Table 4. In the case of PP states of the light sector, $\gamma = 0.07$ GeV is taken as $M^{\eta-\bar{K}^+(0^+(0^{++}))} = 1.0255 GeV$ [33]. $\gamma = 0.05$ GeV is taken for PV states of the light sector by using $M^{\rho-\bar{K}(0^-(1^{+-}))} = 1.266 GeV$ [33]. If we consider $M^{K^*-\bar{K}^*(0^+(0^{++}))} = 1.722 GeV$ [28], we get $\gamma = 0$ GeV for VV states of the light sector. The charge conjugation and parity of the meson–antimeson system are given by $C = (-1)^{L_{12}+S_{12}}$ and $P = P_1 P_2 (-1)^{L_{12}}$ respectively, where $L_{12}$ is the relative orbital momentum and $S_{12}$ is the relative total spin of the system. G-parity is also defined as $G = (-1)^{L_{12}+S_{12}+I}$. We summarize the results of the paper as follows:

(i) $K - \bar{K}$ bound state candidates for $f_0(980)$. The state of $f_0(980)$ is $I^G(J^{PC}) = 0^+(0^{++})$ with mass $0.990 GeV$. Also, the decay of $f_0(980)$ to $K\bar{K}$ has been observed [28]. We have stated that $I^G(J^{PC}) = 0^+(0^{++})$ for $K - \bar{K}$ and obtained 0.993 GeV for this case. Thus we have identified $f_0(980)$ as the dimesonic molecular state $K - \bar{K}$. Rathaud and Rai [33] have reported the mass of this dimesonic system as 0.9768 GeV. Our result (0.993 GeV) is in agreement with experiment (0.990 GeV) and the Rathaud and Rai result [33]. In this case the uncertainty of our result with Ref. [33] is 1.65% and 0.3% in comparison with experiment. They predicted $0^{++}$ and -18.38 MeV for the state and B. E. respectively of $K-\bar{K}$ [33]. We have obtained -1.443 MeV for the B. E., which is close to their result.

(ii) $K-\bar{K}^*$ can be a candidate for $h_1(1380)$. In Ref. [33] Rathaud and Rai suggested that $h_1(1380)$ is a P-V $K-\bar{K}^*$ state. They have reported $M(K-\bar{K}^*)=$ 1.374 GeV [33]. We have obtained $M(K-\bar{K}^*)=1.385$ GeV, which is in agreement with $h_1(1380)$ and Ref. [33].

(iii) $\rho-\bar{\rho}$ and $\rho-\bar{\omega}$ have structures like $f_0(1500)$ and $a_0(1450)$ respectively. The quantum number and mass of $f_0(1500)$ are $I^G(J^{PC})=0^+(0^{++})$ and $M$=1.505 GeV [28]. We have compared $f_0(1500)$ as $\rho-\bar{\rho}$ with a state of $0^+(0^{++})$ and $M$=1.478 GeV. The observed state $a_0(1450)$ is identified as $1^-(0^{++})$ with $M$=1.474 GeV [28]. Its two-photon decay has been observed. We have obtained $M$= 1.426 GeV and $\Gamma_{\gamma\gamma}$=0.6034 KeV for $\rho-\bar{\omega}$ ($1^-(0^{++})$).

(iv) We have calculated the masses of the $D_s^*-\bar{D}_s^*$ ($0^+(2^{++})$, $0^-(1^{+-})$, $0^+(0^{++})$) bound states as 4.208 GeV, 4.219 GeV and 4.225 GeV respectively, which are in agreement with the reported value $4.43\pm 0.16$ GeV by Wang [12]. In Ref. [12], this bound state is a candidate for Y(4140).

(v) In addition to the observed molecular resonances $Z_b(10650)$ and $Z_b(10610)$ with $I^G=1^+$, there should exist two or four molecular bottomonium-like states with quantum numbers like $I^G=1^-$ [34]. In this case we have considered the $B^*-\bar{B}^*$ molecular state with $I^G=1^-$ where we have reported the mass and binding energy of $B^*-\bar{B}^*$ as 10.609 GeV and -41.250 MeV respectively. Rathaud and Rai reported 10.590 GeV and -66.29 MeV for these quantities [32]. We have obtained $\psi(0)=0.388$ GeV$^{3/2}$, which is smaller than $\psi(0)=0.856$ GeV$^{3/2}$ [32]. We have identified the $1^+(1^{+-})$ result in $B^*-\bar{B}^*$ with X(10650), which has quantum numbers $I^G J^P=?^+1^+$ according to the 2015 Review of Particle Physics. Namely, the bottomonium-like resonances $Z_b(10610)$ and $Z_b(10650)$ are respectively $B^*\bar{B}-B\bar{B}^*$ and $B^*\bar{B}^*$ molecules [35]. The molecular interpretation of these states is supported by the recent observation of a high rate of their decay into the corresponding heavy meson pair: $Z_b(10610)\rightarrow B^*\bar{B}(B\bar{B}^*)$, $Z_b(10650)\rightarrow B^*\bar{B}^*$ [36].

(vi) Liu and Zhu concluded that Y(4143) is probably a molecular state $D_s^*-\bar{D}_s^*$ with $J^{PC}$=$0^{++}$ or $2^{++}$ [6, 10]. We have concluded the quantum numbers of the molecular state $D_s^*-\bar{D}_s^*$ are $J^{PC}$=$0^{++}$ or $2^{++}$. We have also concluded that the quantum number of $D_s^*-\bar{D}_s^*$ can be $1^{+-}$.

(vii) The value of the mass of X(3823) is $(3823.1\pm 1.8\pm 0.7)MeV$ [28]. We have identified X(3823) as a $D-\bar{D}_s$ molecular state and reported mass and quantum numbers of this state as 3.828 GeV and $J^{PC}=0^{++}$ respectively. Further, Rathaud and Rai have reported 3.8324 GeV for the mass of this state. They suggested X(3823) as a P-P state $D-\bar{D}_s$ [30].

(viii) $\psi(4040)$ has quantum numbers $I^G J^{PC}=0^-1^{--}$ according to the 2015 Review of Particle Physics. We have identified it as $D^*-\bar{D}^*$ with $0^-(1^{+-})$ and a mass of 4.009 GeV. However, its parity is not in agreement with $\psi(4040)$ but $I$, $G$, $J$ and $C$ are in agreement with $\psi(4040)$. Rathaud and Rai suggested that Y(4008) can be considered as a $D^* - \bar{D}^*$ state [30], where they reported $M(D^* - \bar{D}^*) = 4.0089$ GeV and we have obtained $M(D^* - \bar{D}^*) = 4.009$ GeV.

The $D^*\bar{D}^*$ and $D_s^*\bar{D}_s^*$ bound states can be considered as the resonances X(3915) and Y(4140) respectively. Several authors [6,10,11,30,37] have investigated these states. It is interesting that in the case of a heavy quark system, the small kinetic term yields more possibilities for the formation of molecules [38].

The differences between our results and Ref. [32] for the masses of $B_s^* - \bar{B}_s^*$ in the different states in Table 4 are 28.563 MeV, 13.164 MeV, 14.025 MeV, 13.488 MeV, 7.322 and 5.411 MeV.

We have reported our results for the $\pi^0 - \bar{\pi}^0$ di-mesonic system. However, the obtained B. E. (46.129 MeV) is slightly bigger than the B. E. of Ref. [33] (0.3931 MeV), although the obtained wave function at the origin (0.007 GeV$^{3/2}$) and mass (0.314 GeV) of ours are close to (0.0035 GeV$^{3/2}$) and (0.2703 GeV) respectively [33]. Our two-gamma decay width is 0.0976 keV while the result of Ref. [33] is (0.0091 keV). The reason for the differences may be the different choices of potential models, different approaches to solving the Schrodinger equations, and so on.

We have reported the binding energy of the states. In fact some of these states have not been observed yet. They may be accessible at future experiments like LHCb and the forthcoming BelleII and be confirmed as loosely bound molecular states. Our results provide useful references to explore these in future experiments. The simplicity of our approach has another advantage. We have observed that our calculated masses for the heavier meson-antimeson systems are close to the experimental data, as reported in Tables (3)-(7). We have calculated the contributions of the spin-dependent and OPEP, the binding energy and the wave function at the origin for every system. Di-gamma decay widths are also presented in the final columns of Tables 5-7.

## 4 Conclusions

We have computed the masses, binding energies and wave functions of meson-antimeson systems in the light and heavy meson sectors for s-wave states. We compared the results with the experimentally observed data and predicted theoretically systems which did not include $qq$-structure. Our results are comparable with the available data.

**Acknowledgments**
The authors wish to thank Prof. Eef van Beveren (The Centre for Physics of the University of Coimbra (CFisUC)) for his valuable comments.

Table 1. Masses of mesons (in GeV) [28]

| Meson | $K^+$ | $K^0$ | $\eta$ | $\eta'$ | $\rho$ | $\omega$ | $K^*$ | $\phi$ |
|---|---|---|---|---|---|---|---|---|
| Mass | 0.4936 | 0.4976 | 0.5478 | 0.9577 | 0.7754 | 0.7826 | 0.8959 | 1.0194 |
| Meson | $B^0$ | $B^*$ | $B_s^0$ | $B_s^*$ | $D^\pm$ | $D^*$ | $D_s^\pm$ | $D_s^{*\pm}$ |
| Mass | 5.2795 | 5.3252 | 5.3667 | 5.4154 | 1.8696 | 2.0069 | 1.9683 | 2.1121 |

Table 2. Values of the parameters used in our model for dimesonic systems

| $\dfrac{g_8^2}{4\pi}$ | $M_B$ (GeV) | $\Lambda_Q$ (GeV) | $k$ | $\Lambda_0 (fm^{-1})$ |
|---|---|---|---|---|
| 0.67 | 1 | 0.413 | 0.81 | 2.87 |

Table 3. Masses of heavy charmed meson-antimeson states in GeV ($\gamma$ =0.09 GeV).

| system | $I^G(J^{PC})$ | ours (GeV) | Exp. [28] | others (GeV) | $\psi(0)$ $(GeV^{\frac{3}{2}})$ | B. E. (MeV) | B. E. (MeV) [30] | $<V_{SD}>_{0,0}$ (MeV) | $<V_\pi>_{0,0}$ (MeV) |
|---|---|---|---|---|---|---|---|---|---|
| $D-\bar{D}$ | $0^+(0^{++})$ | 3.729 | - | 3.733 [30], 3.738 [29] | 0.159 | -9.295 | -5.776 | 0 | 0 |
| $D-\bar{D}_s$ | $\dfrac{1}{2}(0^{++})$ | 3.828 | - | 3.832 [30] | 0.162 | -9.421 | -15.95 | 0 | 0 |
| $D-\bar{D}^*$ | $0^-(1^{+-})$ | 3.875 | - | 3.876 [31], 3.871 [30] | 0.163 | -0.731 | -5.600 | 0 | 8.736 |
| $D-\bar{D}^*$ | $1^+(1^{+-})$ | 3.864 | - | 3.871 [30] | 0.163 | -12.381 | -5.600 | 0 | -2.912 |
| $D^*-\bar{D}^*$ | $1^-(2^{++})$ | 4.009 | - | 4.062 [29] | 0.168 | -4.005 | -4.470 | 2.800 | 2.849 |
| $D^*-\bar{D}^*$ | $0^-(1^{+-})$ | 4.009 | $\psi$ (4.040) | 4.0089 [30] | 0.168 | -3.907 | -5.060 | -2.800 | 8.547 |

| system | $I^G(J^{PC})$ | ours | [30] | $\psi(0)$ $(GeV^{\frac{3}{2}})$ | B. E. (MeV) | B. E. (MeV) [30] | $<V_{SD}>_{0,0}$ (MeV) | $<V_\pi>_{0,0}$ (MeV) | |
|---|---|---|---|---|---|---|---|---|---|
| $D^*-\bar{D}^*$ | $0^+(0^{++})$ | 4.015 | - | 4.0083 [30] | 0.168 | 1.839 | -5.658 | -5.600 | 17.094 |
| $D^*-\bar{D}^*$ | $1^-(0^{++})$ | 3.992 | - | 4.0083 [30] | 0.168 | -20.953 | -5.658 | -5.600 | -5.698 |
| $D^*-\bar{D}^*$ | $1^+(1^{+-})$ | 3.998 | - | 4.0089 [30] | 0.168 | -15.304 | -5.060 | -2.800 | -2.849 |
| $D^*-\bar{D}^*$ | $0^+(2^{++})$ | 3.998 | - | 4.0094 [30] | 0.168 | -15.401 | -4.470 | 2.800 | -8.547 |
| $D_s-\bar{D}_s^*$ | $0^-(1^{+-})$ | 4.079 | - | 4.075 [30] | 0.170 | -1.282 | -5.519 | 0 | 8.453 |
| $D^*-\bar{D}_s^*$ | $\frac{1}{2}(0^{++})$ | 4.103 | - | 4.113 [30] | 0.172 | -15.262 | -5.458 | -5.472 | 0 |
| $D_s-\bar{D}_s$ | $0^+(0^{++})$ | 3.927 | - | 3.931 [30] | 0.166 | -9.553 | -5.692 | 0 | 0 |
| $D-\bar{D}_s^*$ | $\frac{1}{2}(1^{+-})$ | 3.972 | - | 3.976 [30] | 0.167 | -9.593 | -5.345 | 0 | 0 |
| $D_s^*-\bar{D}_s^*$ | $0^+(2^{++})$ | 4.208 | - | - | 0.175 | -15.530 | - | 2.675 | -8.270 |
| $D_s^*-\bar{D}_s^*$ | $0^-(1^{+-})$ | 4.219 | - | - | 0.175 | -4.340 | - | -2.675 | 8.270 |
| $D_s^*-\bar{D}_s^*$ | $0^+(0^{++})$ | 4.225 | - | - | 0.175 | 1.254 | - | -5.351 | 16.541 |
| $D_s^*-\bar{D}_s^*$ | $1^-(0^{++})$ | 4.203 | - | - | 0.175 | -20.800 | - | -5.351 | -5.513 |
| $D_s^*-\bar{D}_s^*$ | $1^+(1^{+-})$ | 4.208 | - | - | 0.175 | -15.368 | - | -2.675 | -2.756 |
| $D_s^*-\bar{D}_s^*$ | $1^-(2^{++})$ | 4.219 | - | - | 0.175 | -4.502 | - | 2.675 | 2.756 |

Table 4. Masses of heavy bottom meson-antimeson states in GeV ($\gamma = 0.07$ GeV).

| system | $I^G(J^{PC})$ | ours | [32] | $\psi(0)$ $(GeV^{\frac{3}{2}})$ | B. E. (MeV) | B. E. (MeV) [32] | $<V_{SD}>_{0,0}$ (MeV) | $<V_\pi>_{0,0}$ (MeV) |
|---|---|---|---|---|---|---|---|---|
| $B-\bar{B}$ | $0^+(0^{++})$ | 10.515 | 10.516 | 0.385 | -43.978 | -43.06 | 0 | 0 |
| $B - \bar{B}_s$ | $\frac{1}{2}(0^{++})$ | 10.602 | 10.594 | 0.388 | -44.209 | -51.44 | 0 | 0 |
| $B - \bar{B}^*$ | $0^-(1^{+-})$ | 10.565 | 10.542 | 0.387 | -39.331 | -62.54 | 0 | 4.768 |
| $B_s - \bar{B}^*$ | $\frac{1}{2}(1^{+-})$ | 10.647 | - | 0.390 | -44.333 | - | 0 | 0 |
| $B^* - \bar{B}^*$ | $1^+(1^{+-})$ | 10.603 | 10.586 | 0.388 | -47.194 | -58.27 | -1.386 | -1.585 |
| $B^* - \bar{B}^*$ | $0^+(0^{++})$ | 10.612 | 10.542 | 0.388 | -37.480 | -88.68 | -2.772 | 9.515 |
| $B^* - \bar{B}^*$ | $0^-(1^{+-})$ | 10.609 | 10.567 | 0.388 | -40.851 | -74.84 | -1.386 | 4.757 |
| $B^* - \bar{B}^*$ | $1^-(0^{++})$ | 10.600 | 10.585 | 0.388 | -50.167 | -54.45 | -2.772 | -3.171 |
| $B^* - \bar{B}^*$ | $1^-(2^{++})$ | 10.609 | 10.590 | 0.388 | -41.250 | -66.29 | 1.386 | 1.585 |
| $B_s - \bar{B}_s^*$ | $0^-(1^{+-})$ | 10.742 | 10.727 | 0.393 | -39.848 | -54.27 | 0 | 4.725 |
| $B_s - \bar{B}_s$ | $0^+(0^{++})$ | 10.689 | 10.690 | 0.391 | -44.444 | -43.46 | 0 | 0 |
| $B_s^* - \bar{B}_s^*$ | $0^+(0^{++})$ | 10.792 | 10.752 | 0.394 | -38.027 | -66.59 | -2.749 | 9.428 |
| $B_s^* - \bar{B}_s^*$ | $0^-(1^{+-})$ | 10.789 | 10.771 | 0.394 | -41.366 | -54.53 | -1.374 | 4.714 |
| $B_s^* - \bar{B}_s^*$ | $0^+(2^{++})$ | 10.782 | 10.799 | 0.394 | -48.045 | -34.02 | 1.374 | -4.714 |
| $B_s^* - \bar{B}_s^*$ | $1^-(0^{++})$ | 10.780 | 10.787 | 0.394 | -50.598 | -37.11 | -2.749 | -3.142 |
| $B_s^* - \bar{B}_s^*$ | $1^+(1^{+-})$ | 10.783 | 10.787 | 0.394 | -47.652 | -40.33 | -1.374 | -1.571 |
| $B_s^* - \bar{B}_s^*$ | $1^-(2^{++})$ | 10.789 | 10.787 | 0.394 | -41.759 | -47.17 | 1.374 | 1.571 |
| $B - \bar{B}_s^*$ | $\frac{1}{2}(1^{+-})$ | 10.650 | - | 0.390 | -44.337 | - | 0 | 0 |

Table 5. Masses of light meson-antimeson sector for PP states in GeV ($\gamma = 0.07$ GeV).

| PP states | $I^G(J^{PC})$ | ours | Exp. [28] | others | $\psi(0)$ $(GeV^{\frac{3}{2}})$ | B. E. (MeV) | B. E. (Other) | $<V_{SD}>_{0,0}$ (MeV) | $<V_\pi>_{0,0}$ (MeV) | $\Gamma_{\gamma\gamma}$ (KeV) |
|---|---|---|---|---|---|---|---|---|---|---|
| $\pi^0 - \bar{\pi}^0$ | $0^+(0^{++})$ | 0.314 | - | 0.2703 [33] | 0.007 | 46.129 | 0.3931 [33] | 0 | 0 | 0.0976 |
| $\eta - \bar{\pi}^0$ | $1^-(0^{++})$ | 0.707 | - | 0.6829 [33] | 0.017 | 26.045 | 0.1107 [33] | 0 | 0 | 0.0979 |
| $K^+ - \bar{K}^+$ | $0^+(0^{++})$ | 0.986 | - | 0.9692 [33] | 0.045 | -1.096 | -18.09 [33] | 0 | 0 | 0.3602 |
| $K - \bar{K}$ | $0^+(0^{++})$ | 0.993 | 0.990 | 0.9768 [33] | 0.046 | -1.443 | −20.87 [16] | 0 | 0 | 0.3616 |
| $\eta - \bar{K}$ | $\frac{1}{2}(0^{++})$ | 1.028 | - | 1.0294 [33] | 0.055 | -16.768 | -16.01 [33] | 0 | 0 | 0.4846 |
| $\eta - \bar{\eta}$ | $0^+(0^{++})$ | 1.058 | - | 1.0789 [33] | 0.067 | -36.990 | -26.24 [16] | 0 | 0 | 0.6745 |
| $\eta - \bar{\eta}'$ | $0^+(0^{++})$ | 1.455 | - | 1.458 [16] | 0.087 | -49.850 | -46.75 [16] | 0 | 0 | 0.6070 |

Table 6. Masses of light meson-antimeson sector for PV states in GeV ($\gamma$ =0.05 GeV).

| PV states | $I^G(J^{PC})$ | ours | Exp. [28] | [others] | $\psi(0)$ $(GeV^{\frac{3}{2}})$ | B. E. (MeV) | B. E. (other) | $<V_{SD}>_{0,0}$ (MeV) | $<V_\pi>_{0,0}$ (MeV) | $\Gamma_{\gamma\gamma}$ (KeV) |
|---|---|---|---|---|---|---|---|---|---|---|
| $\eta - \bar{\omega}$ | $0^-(1^{+-})$ | 1.285 | - | 1.330 [33] | 0.080 | -44.410 | −63.22 [16], 0.0504 [33] | 0 | 21.558 | 0.6582 |
| $\eta - \bar{\rho}$ | $1^+(1^{+-})$ | 1.271 | - | 1.259 [16], 1.323 [33] | 0.070 | -52.101 | −63.82 [16], 0.0504 [33] | 0 | -5.588 | 0.5086 |
| $K - \bar{K}^*$ | $0^-(1^{+-})$ | 1.374 | 1.386 | 1.330 [16], 1.383 [33] | 0.061 | -19.016 | -13.60 [33] | 0 | 12.546 | 0.3362 |

| | $I^G(J^{PC})$ | ours | Exp. | [16] | $\psi(0)$ $(GeV^{\frac{3}{2}})$ | B.E. (MeV) | B.E. [16] | $<V_{SD}>_{0,0}$ (MeV) | $<V_\pi>_{0,0}$ (MeV) | $\Gamma_{\gamma\gamma}$ (KeV) |
|---|---|---|---|---|---|---|---|---|---|---|
| $K-\bar{K}^*$ | $1^+(1^{+-})$ | 1.357 | - | 1.330 [16] | 0.061 | -35.745 | -13.60 [33] | 0 | -4.182 | 0.3445 |
| $\eta'-\bar{\omega}$ | $0^-(1^{+-})$ | 1.682 | 1.594 | 1.646 [16] | 0.108 | -57.457 | −93.45 [16] | 0 | 21.075 | 0.6930 |
| $\eta-\bar{\phi}$ | $0^-(1^{+-})$ | 1.516 | - | 1.493 [16] | 0.089 | -50.768 | −73.50 [16] | 0 | 20.124 | 0.5858 |
| $\eta'-\bar{\rho}$ | $1^+(1^{+-})$ | 1.670 | - | 1.639 [16] | 0.094 | -62.498 | −94.25 [16] | 0 | -5.493 | 0.5301 |

Table 7. Masses of light meson-antimeson sector for VV states in GeV ($\gamma = 0$ GeV).

| VV states | $I^G(J^{PC})$ | ours | Exp. [28] | [16] | $\psi(0)$ $(GeV^{\frac{3}{2}})$ | B.E. (MeV) | B.E. [16] | $<V_{SD}>_{0,0}$ (MeV) | $<V_\pi>_{0,0}$ (MeV) | $\Gamma_{\gamma\gamma}$ (KeV) |
|---|---|---|---|---|---|---|---|---|---|---|
| $\rho-\bar{\rho}$ | $0^+(0^{++})$ | 1.478 | 1.505 | 1.489 | 0.075 | -72.121 | −55.39 | -11.277 | 27.055 | 0.4335 |
| $\rho-\bar{\rho}$ | $0^-(1^{+-})$ | 1.470 | - | 1.492 | 0.075 | -80.010 | −55.00 | -5.638 | 13.527 | 0.4382 |
| $\rho-\bar{\rho}$ | $0^+(2^{++})$ | 1.455 | - | 1.500 | 0.075 | -95.788 | −54.10 | 5.638 | -13.527 | 0.4478 |
| $\rho-\bar{\rho}$ | $1^+(1^{+-})$ | 1.452 | - | 1.493 | 0.075 | -98.047 | −54.40 | -5.638 | -4.509 | 0.4492 |
| $\rho-\bar{\rho}$ | $1^-(2^{++})$ | 1.473 | - | 1.499 | 0.075 | -77.751 | −54.70 | 5.638 | 4.509 | 0.4369 |
| $\rho-\bar{\rho}$ | $1^-(0^{++})$ | 1.442 | - | 1.490 | 0.075 | -108.195 | −54.25 | -11.277 | -9.018 | 0.4555 |
| $\omega-\bar{\omega}$ | $0^+(0^{++})$ | 1.461 | - | 1.502 | 0.098 | -103.915 | −55.77 | -22.578 | 43.614 | 0.7629 |
| $\omega-\bar{\omega}$ | $0^-(1^{+-})$ | 1.450 | - | 1.506 | 0.098 | -114.433 | −55.31 | -11.289 | 21.807 | 0.7740 |
| $\omega-\bar{\omega}$ | $0^+(2^{++})$ | 1.429 | - | 1.514 | 0.098 | -135.468 | −54.40 | 11.289 | -21.807 | 0.7970 |
| $\omega-\bar{\omega}$ | $1^-(0^{++})$ | 1.403 | - | - | 0.098 | -162.067 | - | -22.578 | -14.538 | 0.8275 |
| $\omega-\bar{\omega}$ | $1^+(1^{+-})$ | 1.421 | - | - | 0.098 | -143.509 | - | -11.289 | -7.269 | 0.8060 |

| system | $I^G(J^{PC})$ | | | | | | | | |
|---|---|---|---|---|---|---|---|---|---|
| $\omega-\bar{\omega}$ | $1^-(2^{++})$ | 1.458 | - | - | 0.098 | -106.392 | - | 11.289 | 7.269 | 0.7655 |
| $\rho-\bar{\omega}$ | $0^+(0^{++})$ | 1.472 | - | - | 0.085 | -85.679 | - | -15.716 | 34.044 | 0.5668 |
| $\rho-\bar{\omega}$ | $1^-(0^{++})$ | 1.426 | 1.474 | 1.497 | 0.085 | -131.073 | −54.40 | -15.716 | -11.348 | 0.6034 |
| $\rho-\bar{\omega}$ | $1^-(2^{++})$ | 1.467 | - | 1.506 | 0.085 | -90.475 | −54.85 | 7.858 | 5.674 | 0.5705 |
| $\rho-\bar{\omega}$ | $0^-(1^{+-})$ | 1.463 | - | - | 0.085 | -94.843 | - | -7.858 | 17.022 | 0.5739 |
| $\rho-\bar{\omega}$ | $0^+(2^{++})$ | 1.444 | - | - | 0.085 | -113.172 | - | 7.858 | -17.022 | 0.5885 |
| $\rho-\bar{\omega}$ | $1^+(1^{+-})$ | 1.440 | - | - | 0.085 | -117.54 | - | -7.858 | -5.674 | 0.5921 |

Table 8. Excited spectra of heavy charmed meson-antimeson states for $n=1$ in GeV ($\gamma=0.09$GeV).

| system | $I^G(J^{PC})$ | ours (GeV) | $\psi(0)$ $(GeV^{\frac{3}{2}})$ | B. E. (MeV) | $<V_{SD}>_{1,0}$ (MeV) | $<V_\pi>_{1,0}$ (MeV) |
|---|---|---|---|---|---|---|
| $D-\bar{D}$ | $0^+(0^{++})$ | 3.804 | 0.056 | 65.176 | 0 | 0 |
| $D-\bar{D}_s$ | $\frac{1}{2}(0^{++})$ | 3.903 | 0.057 | 65.144 | 0 | 0 |
| $D-\bar{D}^*$ | $0^-(1^{+-})$ | 3.942 | 0.057 | 66.197 | 0 | 1.064 |
| $D-\bar{D}^*$ | $1^+(1^{+-})$ | 3.941 | 0.057 | 64.778 | 0 | -0.354 |
| $D^*-\bar{D}^*$ | $1^-(2^{++})$ | 4.079 | 0.059 | 65.783 | 0.350 | 0.346 |
| $D^*-\bar{D}^*$ | $0^-(1^{+-})$ | 4.079 | 0.059 | 65.776 | -0.350 | 1.040 |
| $D^*-\bar{D}^*$ | $0^+(0^{++})$ | 4.080 | 0.059 | 66.466 | -0.700 | 2.080 |
| $D^*-\bar{D}^*$ | $1^-(0^{++})$ | 4.077 | 0.059 | 63.692 | -0.700 | -0.693 |
| $D^*-\bar{D}^*$ | $1^+(1^{+-})$ | 4.078 | 0.059 | 64.389 | -0.350 | -0.346 |

| System | $I^G(J^{PC})$ | | | | | |
|---|---|---|---|---|---|---|
| $D^* - \bar{D}^*$ | $0^+(2^{++})$ | 4.078 | 0.059 | 64.396 | 0.350 | -1.040 |
| $D_s - \bar{D}_s^*$ | $0^-(1^{+-})$ | 4.146 | 0.060 | 66.094 | 0 | 1.028 |
| $D^* - \bar{D}_s^*$ | $\frac{1}{2}(0^{++})$ | 4.183 | 0.060 | 64.368 | -0.684 | 0 |
| $D_s - \bar{D}_s$ | $0^+(0^{++})$ | 4.001 | 0.058 | 65.111 | 0 | 0 |
| $D - \bar{D}_s^*$ | $\frac{1}{2}(1^{+-})$ | 4.046 | 0.059 | 65.101 | 0 | 0 |
| $D_s^* - \bar{D}_s^*$ | $0^+(2^{++})$ | 4.288 | 0.062 | 64.345 | 0.334 | -1.005 |
| $D_s^* - \bar{D}_s^*$ | $0^-(1^{+-})$ | 4.289 | 0.062 | 65.687 | -0.334 | 1.005 |
| $D_s^* - \bar{D}_s^*$ | $0^+(0^{++})$ | 4.290 | 0.062 | 66.357 | -0.668 | 2.010 |
| $D_s^* - \bar{D}_s^*$ | $1^-(0^{++})$ | 4.287 | 0.062 | 63.677 | -0.668 | -0.670 |
| $D_s^* - \bar{D}_s^*$ | $1^+(1^{+-})$ | 4.288 | 0.062 | 64.346 | -0.334 | -0.335 |
| $D_s^* - \bar{D}_s^*$ | $1^-(2^{++})$ | 4.289 | 0.062 | 65.685 | 0.334 | 0.335 |